\documentclass[12pt,preprint]{aastex}
\begin{document}
\title{Lane-Emden Equation: perturbation method}
\author{Zakir F. Seidov}
\affil{College of Judea and Samaria,Israel}
\begin{abstract}
The perturbation method is applied to\,numerical solution of the Lane-Emden
Equation (LEE)of arbitrary index $n$, and the global parameters of polytropes
are found as function of polytropic index $n$.\end{abstract}
\section{Introduction}The basic differential equation of internal
structure of stars, the Lane-Emden equation (hereafter LEE) of index $n$,
is solved analytically only for three values of $n=0, 1, 5$.
For other values of  $n$, LEE can be solved only\,numerically.
Recently, using the perturbation method, SK ( Seidov \& Kuzakhmedov, 1978 (SK78))
had presented the new {\it analytical} solutions of the LEE for index $n$ only
slightly differing from 0, 1, and 5, see also
\citet{Seid78a,Seid78b,Seid79a,Seid79b,Seid04,
Jabb84,Caim87,Hord87,Hord90,MoRy01}.\\
In this paper I  present the {\it\,numerical} perturbation method for solving
the LEE, see also \citet{Seid04}. \section{Basic equation}
The  basic equation is LEE of index $n$ : \begin{equation}
LE=\frac{1}{x^{2}}\frac{d}{d\, x}\left( x^{2}\,\frac{d}{d\, x}\right);
\quad LE[y]=-y^{n};\quad y(0)=1,\quad y^{'}(0)=0,\label{LEE} \end{equation}
where we introduced the Lane-Emden differental operator {\it LE}.\\
We look for solution $y(x)$ in the interval $[0,X]$ such that $y(X)=0$.\\
Three classical analytical solutions of equation
(\ref{LEE}) are (see e.g. \citep{Chan57}):
\begin{equation}\,n=0,\quad y=1-\frac{1}{6}\,x^{2},\quad X=\sqrt{6},\quad
\mu =2\,\sqrt{6},\quad \rho_c/\rho_m=1; \label{LEE0}\end{equation}
\begin{equation}\,n=1,\quad y=\frac{\sin x}{x},\quad X=\pi,\quad
\mu =\pi,\quad \rho_c/\rho_m=\pi^{2}/3; \label{LEE1}\end{equation}
\begin{equation}\,n=5,\quad y=(1+\frac{1}{3}\,x^{2})^{-1/2},
\quad X\rightarrow \infty ,\quad \mu =\sqrt{3},\quad
\rho_c/\rho_m\rightarrow\infty. \label{LEE5}\end{equation}
In these equations, $\mu=-X^{2}y^{'}(X),\quad \rho_c/\rho_m=X^{3}/3\,\mu$; $X$, $\mu$
are dimensionless radius and mass, and $\rho_c/\rho_m$ is the central-to-mean density
ratio. \section{The perturbation method}\label{pert}
Consider equation (\ref{LEE}) as ODE depending on parameter $n$, then assuming
$n=n_{0}+\delta$ with $\delta$ as a small parameter, $\delta\ll\,n_{0}$,
(or $\delta\ll 1$, if $n_{0}=0$) we expand the r.s. of equation (\ref{LEE})
to the second order of $\delta$ (for the sake of brevity we omit index at $n$):
\begin{eqnarray} y = y_{0} + \delta\,  y_{1}+ \delta^{2}\,  y_{2};\quad
y_{0}^{n+\delta}= h_{0}+\delta\,h_{0}+\delta^{2}\,h_{1}; \nonumber \\
 h_{0}=y_{0}^{n};\quad h_{1}= \,n\, y_{0}^{n-1}\,y_{1} + y_{0}^{n}\,\ln y_{0}; \nonumber \\
h_{2}={1\over2}\,n\,(n-1)\, y_{0}^{n-2}\, y_{1}^{2} + y_{0}^{n-1} \,(y_{1}
+n\,y_{2} +\,n\,y_{1}\,\ln y_{0}) + {1\over 2}\,y_{0}^{n}\,\ln^{2} y_{0}.
\label{rsn}\end{eqnarray} From  equations (\ref{LEE},\ref{rsn}) we have
three coupled ODEs for three functions $y_{0},\, y_{1},\, y_{2}$: \begin{equation}
LE[ y_{0}] =- h_{0},\quad y_{0}(0)=1,\quad y_{0}^{'}(0)=0; \label{eq0} \end{equation}
 \begin{equation} LE[ y_{1}] =- h_{1},\quad y_{1}(0)= y_{1}^{'}(0)=0;
 \label{eq1} \end{equation}  \begin{equation}
LE[ y_{2}] = - h_{2},\quad y_{2}(0)= y_{2}^{'}(0)=0. \label{eq2} \end{equation}
Initial conditions in equations (\ref{eq0},\ref{eq1},\ref{eq2})
are defined by the form of series expansion of the solution
of LEE of arbitrary $n$ at $x=0$ (see further, formulas (\ref{ser00} - \ref{ccii}):
\begin{equation} y=1-\frac{1}{6}\,x^{2}+\frac{n}{5!}\,x^{4}
+\ldots. \label{ys} \end{equation}
Writing $n=n_{0}+\delta$, expanding equation (\ref{ys}) to the second order
of $\delta$, we have the series expansions for functions $y_{1},\,y_{2}$ at $x=0$:
\begin{eqnarray}\label{y1ser}
y_{1}=\frac{1}{5!}\,x^{4}+{5 - 16\,n\over 3\cdot 7!}\,x^{6} \ldots , \end{eqnarray}
\begin{eqnarray}\label{y2ser} y_{2}=-\frac{8}{3\cdot 7!}\,x^{6}+{183\,(-1
 + 2\,n)\over 9\cdot 9!}\,x^{8} \ldots . \end{eqnarray}
Note that the series expansion for $y_{1}$ and  $  y_{2}$
were given in \cite{Seid04} only for case $n=0$.\\
Before solving eqs  (\ref{eq0},\ref{eq1},\ref{eq2}), I'd like to mention that
the validity of the approach used is discussed briefly in
SK78, and in \cite{Seid04}. For $n=0,\,1,\,\mbox{and }\, 5$ the
 {\it analytical} solutions for functions  $y_{1}$
 were presented in SK78; also the second approximation
at the case of $n=0$ is partly given in \cite{Seid04}. Here we
  solve\,{\it numerically} equations (\ref{eq0} - \ref{eq2}).
  We remind that $y_{0}$ is the "basic" solution of LEE and
we refer to $y_{1}$ and $y_{2}$ as the "perturbed" LEE solutions
of the first and second order. I widely used {\it MATEMATICA}'s \citep{Wolf99}
function\, {\it NDSolve} with suitable options.
\section{System of ODEs}\label{system}
For a better numerical accuracy, we introduce three additional functions,
$z_{0}$,  $z_{1}$, and $z_{2}$: \begin{equation}\label{z0z1}
  z_{0}=-x^{2}\,y_{0}^{'},\quad z_{1}=-x^{2}\,y_{1}^{'},
  \quad z_{2}=-x^{2}\,y_{2}^{'},\end{equation}
and rewrite the three equations (\ref{eq0}-\ref{eq2}) as a system of
six differential equations: \begin{equation}
y_{0}^{'}=-z_{0}/x^{2},\,\quad z_{0}^{'}=-x^{2}\,h_{0},
\label{sys4a}\end{equation}. \begin{equation}
y_{1}^{'}=-z_{1}/x^{2},\,\quad z_{1}^{'}=-x^{2}\,h_{1}. \label{sys4b}\end{equation},
 \begin{equation}
y_{2}^{'}=-z_{2}/x^{2},\,\quad z_{2}^{'}=-x^{2}\,h_{2}. \label{sys4c}\end{equation}.
\section{Series solution of zero-th order}\label{seryz0}
Again, for a larger\,numerical accuracy, we use the series solution of LEE.
In \cite{SK77}, the method of accurate series solution of LEE is given. Using
the formulas of \cite{SK77}, we present solution of LEE (\ref{sys4a})
at $x=0$ in the form \begin{eqnarray}
y_{0}=1+\sum_{i=1}^{i=12}\,a_{i}\,x^{2\,i}, \quad
z_{0}=-2\,\sum_{i=1}^{i=12}\,i\,a_{i}\,x^{2\,i+1},\label{ser00} \end{eqnarray}
with coefficients $a_{i}$ as follows: \begin{eqnarray}
a_{1} = -{1\over 6}; \quad a_{2} = {n \over 5!}; \quad
 a_{3} = {n\,(5 - 8 \,n)\over 3 \cdot 7!};\quad
 a_{4} = {n\,(70 - 183\,n + 122\,n^2)\over 9\cdot 9!};\nonumber \\
 a_{5} = {n (3150 - 10805\,n + 12642\,n^2 - 5032\,n^3)\over 45\cdot 11!}; \nonumber \\
a_{6} = {n\,(138600 - 574850\,n + 915935\,n^2 - 663166\,n^3
 + 183616\,n^4)\over 135\cdot 13!}; \nonumber \\
  a_{7}=n\,(21021000-101038350\,n+199037015\,n^2 \nonumber \\
-200573786\,n^3 + 103178392\,n^4 - 21625216\,n^5)/ 945 \cdot 15!; \nonumber \\
 a_{8} =n\, (1891890000 - 10267435500\,n+23780949500\,n^2-30057075285\,n^3 \nonumber \\
 +21827357636\,n^4-8618115372\,n^5+1442431856\,n^6)/2835\cdot 17!; \nonumber \\
a_{9}=n\, (675404730000 - 4066235428500\,n + 10740122081500\,n^2 \nonumber \\
-16120795594195\,n^3 + 14830640277988\,n^4 - 8348507232868\,n^5 \nonumber \\
          + 2657923739344\,n^6 - 368552598784\,n^7)/25515\cdot 19!;\nonumber \\
 a_{10}=n\, (171102531600000-1128186384570000\,n \nonumber \\
 + 3329284073314500\, {n^2}-5740042719521900\, {n^3} \nonumber \\+
6317195348852735\, {n^4}-4538114873629364\, {n^5} \nonumber \\+
2074925918891156\, {n^6}-551199819782480\, {n^7} \nonumber \\+
65035924972928\, {n^8})\big/127575\cdot 21!;  \nonumber \\
a_{11}=n\, (118574054398800000-847953056599110000\,n \nonumber \\+
  2754994980587692500\, {n^2}-5335484162711174500\, {n^3} \nonumber \\+
  6782008348777403475\, {n^4}-5860922969087284308\, {n^5} \nonumber \\+
  3438918097715059380\, {n^6}-1319254687791147504\, {n^7} \nonumber \\+
  299840088682556928\, {n^8}-30720693974199296\, {n^9})\big/ 1403325 \cdot 23!; \nonumber \\
 a_{12}=n\, (27272032511724000000
  -209899877314257900000\,n+742585473289204545000\, {n^2}\nonumber \\
  -  1589853990586539282500\, {n^3}+2279636465710370388750\, {n^4}
  -  2285217511971127632065\, {n^5} \nonumber \\ +1620103707989338077938\, {n^6}-
  801095938682391176900\, {n^7}+264081052577164986584\, {n^8}\nonumber \\-
  52342890902954850528\, {n^9} +4731477379473053696\, {n^{10}})\big/
 4209975 \cdot 25!. \label{aaii} \end{eqnarray}
We remind that the general recurrence relation for coefficient $a_i$ is
(\cite{SK77}): \begin{equation}\label{recur}
 a_{m+1}={1\over m(m+1)(2 m+3)}\sum_{i=1}^{i=m}\,(i\,n+i-m)(m+1-i)[3+2(m-i)]a_i\,
 a_{m+1-i}. \end{equation}\section{Series solution of the first order}\label{seryz1}
Writing $n=n_{0}+\delta$, expanding equation (\ref{aaii}) to the {\it first} order
of $\delta$, we have, from (\ref{ser00}),  the series expansions for
solutions $y_{1},\,z_{1}$ of the equations (\ref{sys4b}) at $x=0$:
\begin{eqnarray} y_{1}=1+\sum_{i=2}^{i=12}\,b_{i}\,x^{2\,i},\quad
z_{1}=-2\,\sum_{i=2}^{i=12}\,i\,b_{i}\,x^{2\,i+1},\label{ser11} \end{eqnarray}
with coefficients $b_{i}$ as follows: \begin{eqnarray}
b_2 = 1/5!; \quad  b_3 = (5 - 16\,n)/3\cdot 7!; \quad
b_4 = (70 - 366\,n + 366\,n^2)/9\cdot 9!; \nonumber \\
b_5 = (3150 - 21610\,n + 37926\,n^2 - 20128\,n^3)/45\cdot 11!; \nonumber \\
b_6 = (138600 - 1149700\,n + 2747805\,n^2 - 2652664\,n^3 + 918080\,n^4)/
 135\cdot 13!;  \nonumber \\
b_7 = (21021000 - 202076700\,n + 597111045\,n^2 - 802295144\,n^3\nonumber \\ +
              515891960\,n^4 - 129751296\,n^5)/945\cdot 15!; \nonumber \\
b_8 = 4(472972500 - 5133717750\,n + 17835712125\,n^2 - 30057075285\,n^3\nonumber \\ +
              27284197045\,n^4 - 12927173058\,n^5 + 2524255748\,n^6)/
          2835\cdot 17!; \nonumber \\
          b_9 = 4(168851182500 - 2033117714250\,n + 8055091561125\,n^2 \nonumber  \\-
            16120795594195\,n^3 + 18538300347485\,n^4\nonumber \\ - 12522760849302\,n^5 +
            4651366543852\,n^6 - 737105197568\,n^7)/25515\cdot 19!; \nonumber \\
b_{10} = (171102531600000 - 2256372769140000\,n + 9987852219943500\,n^2 \nonumber \\ -
            22960170878087600\,n^3 + 31585976744263675\,n^4 \nonumber \\ -
            27228689241776184\,n^5 + 14524481432238092\,n^6 \nonumber \\ -
            4409598558259840\,n^7 + 585323324756352\,n^8)/127575\cdot 21!; \nonumber \\
b_{11} = (118574054398800000 - 1695906113198220000\,n \nonumber \\+
            8264984941763077500\,n^2 - 21341936650844698000\,n^3\nonumber \\ +
            33910041743887017375\,n^4 - 35165537814523705848\,n^5\nonumber \\ +
            24072426684005415660\,n^6 - 10554037502329180032\,n^7 \nonumber \\ +
            2698560798143012352\,n^8 - 307206939741992960\,n^9)/
         1403325 \cdot 23!;\nonumber \\
b_ {12} = 2 \,(13636016255862000000 - 209899877314257900000\,n\nonumber \\ +
                1113878209933806817500\,n^2 - 3179707981173078565000\,n^3\nonumber \\ +
                5699091164275925971875\,n^4 - 6855652535913382896195\,n^5\nonumber  \\+
                5670362977962683272783\,n^6 - 3204383754729564707600\,n^7\nonumber  \\+
                1188364736597242439628\,n^8 - 261714454514774252640\,n^9\nonumber \\ +
 26023125587101795328\,n^{10})/\ 4209975\cdot\ 25!. \label{bbii} \end{eqnarray}
\section{Series solution of the second order}\label{seryz2}
Again, as in section {\ref{seryz1},
writing $n=n_{0}+\delta$, expanding equation (\ref{aaii}) to the {\it second} order
of $\delta$, we have get  the series expansions for  solutions $y_{2},\,z_{2}$ of
  the equations (\ref{sys4c}) at $x=0$: \begin{eqnarray}
y_{2}=1+\sum_{i=3}^{i=12}\,c_{i}\,x^{2\,i},\quad
z_{2}=-2\,\sum_{i=3}^{i=12}\,i\,c_{i}\,x^{2\,i+1},\label{ser22} \end{eqnarray}
with coefficients $c_{i}$ as follows: \begin{eqnarray} c_3 = -1/3\cdot7!;\quad
c_4 = 61\, (-1 + 2\, n)/3\cdot 9!;\nonumber \\
c_5 = (-10805 + 37926\, n - 30192\, n^2)/45\cdot 11!; \nonumber \\
c_6 = (-574850 + 2747805\, n - 3978996\, n^2 + 1836160\, n^3)/135\cdot 13!;\nonumber \\
c_7 = (-101038350 + 597111045\, n - 1203442716\, n^2 \nonumber \\ +
1031783920\, n^3 - 324378240\, n^4)/ 945\cdot15!;\nonumber \\
c_8 = 2 \,(-5133717750 + 35671424250\, n - 90171225855\, n^2 \nonumber \\+
109136788180\, n^3 - 64635865290\, n^4 + 15145534488\, n^5))/ 2835\cdot17!;\nonumber \\
c_9 = 2 \,(-2033117714250 + 16110183122250\, n - 48362386782585\, n^2 \nonumber \\ +
74153201389940\, n^3 - 62613804246510\, n^4\nonumber \\ +
27908199263112\, n^5 - 5159736382976\, n^6)/ 25515\cdot19!;\nonumber \\
c_{10} = 2 \,(-564093192285000 + 4993926109971750\, n \nonumber \\-
17220128158565700\, n^2 + 31585976744263675\, n^3 \nonumber \\-
34035861552220230\, n^4 + 21786722148357138\, n^5 \nonumber \\-
7716797476954720\, n^6 + 1170646649512704\, n^7)/127575\cdot21!; \nonumber \\
c_ {11} = 2(-141325509433185000 + 1377497490293846250\, n  \nonumber \\ -
5335484162711174500\, n^2 + 11303347247962339125\, n^3 \nonumber \\ -
14652307422718210770\, n^4 + 12036213342002707830\, n^5  \nonumber \\ -
6156521876358688352\, n^6 + 1799040532095341568\, n^7  \nonumber \\ -
230405204806494720\, n^8)/ 467775\cdot 23!; \nonumber \\
c_ {12} =(-209899877314257900000 + 2227756419867613635000\, n \nonumber \\
- 9539123943519235695000\,n^2 + 22796364657103703887500\,n^3\nonumber \\
- 34278262679566914480975\,n^4 + 34022177867776099636698\,n^5\nonumber  \\
- 22430686283106952953200\,n^6 +9506917892777939517024\,n^7\nonumber  \\
-2355430090632968273760\,n^8 +260231255871017953280\,n^9)/\ 4209975\cdot\ 25!.
\label{ccii} \end{eqnarray} \section{Cases with analytical solutions}\label{compar}
We first consider three cases, for which the zero-th and first approximations have  exact
analytical solutions. \subsection{n=5 case}\label{n5case}
We first mention that in this case coefficients $a_{i}$ can be found
from the very simple recurrence relation:  \begin{eqnarray}n=5:\quad
 a_{m}={1\over 3\,m}\,\left({3\over 2}-m\right)\,a_{m-1},\quad m\geq 1,
 \quad a_{0}=1, \label{recrel5} \end{eqnarray}  which can be used e.g.
 for checking formulas (\ref{ser00},\,\ref{aaii}). Then, we calculate, for $n=5$,
values of  $y_{0}$ and $z_{0}$ according to (\ref{ser00})
and (\ref{aaii}) at point $x_{i}=1/100$. With using {\it MATHEMATICA} it can be done
 {\it exactly} (without any {\it\,numerical error}). Then we compare these values
 with exact analytical values of  $y_{0}$ and $z_{0}$ at $x_{i}=1/100$
 (see (\ref{LEE5})): \begin{eqnarray}\label{comp5}
 { y_{0}(x_{i})_{15,16}\over (1+x_{i}^{2}/3) ^{-1/2}}-1= 9.72065376\cdot 10^{-60},\nonumber \\
 {z_{0}(x_{i})_{15,16}\over (x_{i}^{3}/3)(1+x_{i}^{2}/3)^{-3/2}}-1=
  -2.5273216\cdot 10^{-60}. \end{eqnarray}
  Hence we have some 60 digits of accuracy in the initial values for LEE
  (\ref{sys4a}). Now we solve the system  (\ref{sys4a}-\ref{sys4c})
  numerically using {\it NDSolve} procedure of {\it MATHEMATICA},
  from $x=x_{i}=1/100$ till $x=x_{f}=10$ with options: {Method $\rightarrow$ RungeKutta,
  AccuracyGoal $\rightarrow$ Infinity,  PrecisionGoal $\rightarrow$ 32,
  WorkingPrecision $\rightarrow$ 24,  MaxSteps $\rightarrow$ 50000}, \
  and compare the\,numerical values with values from analytical expressions
  when available \begin{equation}\label{n5sol} \begin{array}{cc}
-------------- & ------------\\
 \,n=5 & x_{f}=10 \\ -------------- & ------------\\
    y_{0}(x_{f}),\mbox{NDSolve} &  0.17066\, 40371\, 96572\,26797\,786 \\
  y_{0}(x_{f}),\mbox{Analytic} & 0.17066\, 40371\, 96572\,28860\,143\\
  -------------- & ------------\\
  z_{0}(x_{f}),\mbox{NDSolve} &  1.65693\, 23999\, 66721\,38805\,555\\
   z_{0}(x_{f}),\mbox{Analytic} & 1.65693\, 23999\, 66721\,24855\,754\\
   -------------- & ------------\\
    y_{1}(x_{f}),\mbox{NDSolve} &  0.096926\, 10440\, 49489\, 40514\, 448 \\
  y_{1}(x_{f}),\mbox{Analytic} & 0.096926\, 10440\, 49489\, 36327\, 188\\
  -------------- & ------------\\
  z_{1}(x_{f}),\mbox{NDSolve} &  -0.12538\, 65905\, 38367\, 83413\, 8922 \\
   z_{1}(x_{f}),\mbox{Analytic} & -0.12538\, 65905\, 38367\, 81528\, 4516\\
   -------------- & ------------\\
   y_{2}(x_{f}),\mbox{NDSolve} &  -0.0131\,49071\,88483\,68785\\
   -------------- & ------------\\
   z_{2}(x_{f}),\mbox{NDSolve} & 0.0001\,04233\,84608\,82482.\\
   -------------- & ------------ \end{array} \end{equation}
Final results are of 15-16 digits accuracy. Here "analytical" values for $y1$ and
$z1$ are given according to analytical solution SK78:
\begin{eqnarray} \label{n5anall1} n=5: \quad
\nu = \arctan ( {x\over \surd 3} ), \quad \nonumber \\
y_{1}={1\over 48\,\sin (\nu)}\, \left(\sin (2\nu) -{5\over 4}\,\sin (4\nu)  +
 3\,\nu\, \cos (4\nu)- 3\,\left[ 2\, \sin (2\nu) + \sin (4\nu) \right]\,
 \ln [\cos (\nu) ]\,\right),\nonumber \\ z_{1}=-x^{2}{d\,y_{1}\over d\,x}.\quad\end{eqnarray}
\subsection{n=1 case}\label{n1case} In the case of $n=1$ we first mention that coefficients $a_{i}$ are of very simple
form: \begin{equation} n=1:\quad a_{i}={1\over (2\,i+1)!}, \label{n1aii}\end{equation}
 which can be used e.g. for checking formulas (\ref{ser00},\,\ref{aaii}).
 Then, we calculate, for $n=1$, values of  $y_{0}$ and $z_{0}$ according to (\ref{ser00})
and (\ref{aaii}) at point $x_{i}=1/100$, and compare these values
 with exact analytical values of  $y_{0}$ and $z_{0}$ at $x_{i}=1/100$:
 \begin{eqnarray}\label{comp1}
 { y_{0}(x_{i})_{16,17}\over\sin(x_{i})/x_{i}}-1= 9.183841796\cdot 10^{-81},\nonumber \\
 { z_{0}(x_{i})_{16,17}\over \sin(x_{i})-x_{i}\,\cos(x_{i})}-1=
-2.3877590\cdot 10^{-81}. \end{eqnarray}
  Now we have some 80 digits of accuracy in the initial values for LEE
  (\ref{sys4a}). Then we solve the system  (\ref{sys4a} - \ref{sys4c})\,numerically using {\it
   NDSolve} procedure of {\it MATHEMATICA}, from $x=x_{i}=1/100$ till $x=x_{f}=\pi$
  with the same options as for case $n=5$, section \ref{n5case}, and compare the
  numerical values with values from analytical expressions when available
  \begin{equation}\label{n1sol} \begin{array}{cc} \,n=1 & X=\pi \\
-------------- & ------------\\
    y_{0}(X),\mbox{NDSolve} &  -6.552265490 \cdot 10^{-19} \\
  y_{0}(X),\mbox{Analytic} & 0\\ -------------- & ------------\\
  z_{0}(X),\mbox{NDSolve} &   3.14159\,26535\,89793\,22713\,584 \\
   z_{0}(X),\mbox{Analytic},\,\pi  & 3.14159\,26535\,89793\,23846\,264\\
-------------- & ------------\\
   y_{1}(X),\mbox{NDSolve} &    0.28179\,14499\,02207\,82015\,93981\\
     y_{1}(X),\mbox{Analytic} &\\
     {1\over 4\pi}\,\mbox{Si}(2\pi)+{1\over 2}\ln (2\pi)-{3\over 4} &
       0.28179\,14499\,02207\,82012\,97922\\ -------------- & ------------\\
     z_{1}(X),\mbox{NDSolve} & -1.02925\,45490\,49509\,89738\,663\\
    z_{1}(X),\mbox{Analytic}  &  \\ {1\over 4}\mbox{Si}(2\pi)+{\pi\over 4}\,[
    \mbox{Ci}(2\pi) +\ln (2\pi)-3-\mbox{EulerGamma}] &-1.02925\,45490\,49509\,87940\,655\\
-------------- & ------------\\  y_{2}(X),\mbox{NDSolve} &  -0.09455\,03188\,95873\,420\\
    y_{2}(X),\mbox{Analytic}& -\\ -------------- & ------------\\
    z_{2}(X),\mbox{NDSolve} &  -0.81171\,80531\,86985\,23026\\
     z_{2}(X),\mbox{Analytic} &  -\\   -------------- & ------------
     \end{array} \end{equation}
Again, as $n=5$ case, final results are of 15-16 digits accuracy, and though there
is\,no guarantee that the similar accuracy will be in other\,numerical results,
still we will use the same set of options for solving other cases of $n$.
\subsection{n=0 case}\label{n0case} In the case of $n=0$ we first mention
that coefficients $a_{i}$ are of very simple form:
\begin{equation} n=0:\quad a_{1}=-{1\over 6},\quad  a_{i>1}=0;\label{n0aii}\end{equation}
while coefficients $b_{i}$ can be found from the series expansion of
 the known function $y_{1}$ (SK78): \begin{equation}
n=0:\quad y_{1}(x)=  \frac{5\,x^2}{18}-4 +\left( 2+y_{0} \right) \,
   \ln({y_0})+ 4\,\frac {\surd 6}{x}\,\mbox{arctanh}\,(\frac{x}{\surd 6}),
     \label{y1n0} \end{equation} with (see equation (\ref{LEE0})):
$$ n=0:\quad y_{0}=1-{1\over 6}\,x^{2}, $$ and
\begin{equation}y_{1}(\surd 6)=  4 \ln2 -{7\over 3}.\label{y1sq6} \end{equation}
 Then we solve, for $n=0$, the system  (\ref{sys4a}-\ref{sys4c}) numerically using {\it
   NDSolve} procedure of {\it MATHEMATICA}, from $x=x_{i}=1/100$ till $x=X=\surd 6$
  with the same options as for cases $n=5,\,1$, and compare the numerical values with
 values from analytical expressions  when available:
\begin{equation}\label{n0sol} \begin{array}{cc} \,n=0 & X=\surd 6 \\
-------------- & ------------\\
    y_{0}(X),\mbox{NDSolve} &   1.530418314373 \cdot 10^{-31} \\
  y_{0}(X),\mbox{Analytic} & 0\\  -------------- & ------------\\
  z_{0}(X),\mbox{NDSolve} &  4.89897\,94855\,66356\,19639\,45681\,49410\,81340 \\
   z_{0}(X),\mbox{Analytic}=2\,\surd 6  & 4.89897\,94855\,66356\,19639\,45681\,49411\,78278\\
-------------- & ------------\\
   y_{1}(X),\mbox{NDSolve} &   0.43925\,53889\,06447\,90287 \\
     y_{1}(X),\mbox{Analytic}=4\,\ln2 -{7\over 3} & 0.43925\,53889\,06447\,90433\\
-------------- & ------------\\  z_{1}(X),\mbox{NDSolve} & -6.27251\,76587\,60946\,466\\
    z_{1}(X),\mbox{Analytic}={4\over 3}\surd 6\,(3\,\ln 2 -4) &
    -6.27251\,76587\,60954\,355\\ -------------- & ------------\\
   y_{2}(X),\mbox{NDSolve} &  -0.41351\,38937\,125\\  y_{2}(X),\mbox{Analytic})=& \\
={1\over 9} (413 - 21\, \pi^{2} -402\, \ln 2 +144\, \ln^{2}2) & -0.41351\,38893\,059\\
   -------------- & ------------\\     z_{2}(X),\mbox{NDSolve} &  -2.59850 \cdot 10^{12}\\
     z_{2}(X),\mbox{Analytic} &  \infty\\ -------------- & ------------
     \end{array} \end{equation} We see that numerical accuracy is drastically decreasing
 for the  larger degrees of perturbation: while $y_{0}(X)$, and $z_{0}(X)$ are
 calculated with some 30-31 correct digits, $y_{1}(X)$, and $z_{1}(X)$ are calculated with
15-16 correct digits, and $y_{2}(X)$, and $z_{2}(X)$ only with  8-9 correct digits.
 \section{Non-analytical cases}\label{nonanal}
For other values of $n$, there is no analytic solution even for
 $y_{0}$ and there is no possibility to check numerical results.\\
 Moreover we should calculate numerically LEE to find accurate values of the
 first zero $X$.
  Also we give comparison of our numerical values of $X$ and $z_{0}(X)$
 with two most accurate results known in literature (\cite{Jabb84},\cite{Hord90}).
 \subsection{n=1/2 case}\label{n12case}
 \begin{equation}\label{n12sol} \begin{array}{cl}
 -------------- & ------------\\ \,n=1/2 &  \\ -------------- & ------------\\
{X},\mbox{NDSolve}& 2.75269\,80540\,64991\\
{X},{\mbox{\cite{Jabb84}}}& 2.75269\,80541\\
{X},{\mbox{\cite{Hord90}}}& 2.75269\,8\\ -------------- & ------------\\
    y_{0}(X),\mbox{NDSolve} &   -1.57415358 \cdot 10^{-15} \\
  y_{0}(X),\mbox{Analytic} & \mbox{  0}\\  -------------- & ------------\\
  z_{0}(X), \mbox{NDSolve} &  3.78865\,11848\,84005\,74354\,209 \\
   z_{0}(X),{\mbox{\cite{Hord90}}}  &3.78865\,11652\,97791\\
-------------- & ------------\\
   y_{1}(X),\mbox{NDSolve} &   0.34123\,20401\,31647\,34451 \\
    -------------- & ------------\\
     z_{1}(X),\mbox{NDSolve} & -1.62970\,77128\,59769\,520\\
   -------------- & ------------\\
   y_{2}(X),\mbox{NDSolve} & -0.14682\,13064\,84602\,959\\
       -------------- & ------------\\
    z_{2}(X),\mbox{NDSolve} &  -3.71857\,97123\,08979\,172\\
         -------------- & ------------  \end{array} \end{equation}
\subsection{n=3/2 case}\label{n32case}
 \begin{equation}\label{n32sol} \begin{array}{cl}
 -------------- & ------------\\ \,n=3/2 &  \\ -------------- & ------------\\
{X},\mbox{NDSolve}& 3.65375\,37362\,19119 \\
{X},{\mbox{\cite{Jabb84}}}& 3.65375\,37362 \\
{X},{\mbox{\cite{Hord90}}}& 3.65375\,4\\ -------------- & ------------\\
    y_{0}(X),\mbox{NDSolve} &   6.8926652958 \cdot 10^{-16} \\
  y_{0}(X),\mbox{Analytic} &  \mbox{ 0}\\ -------------- & ------------\\
  z_{0}(X), \mbox{NDSolve} &  2.71405\,51201\,08646 \\
   z_{0}(X),{\mbox{\cite{Hord90}}}  &2.71405\,57437\\
-------------- & ------------\\
   y_{1}(X),\mbox{NDSolve} &   0.24067\,09191\,40952\,32 \\
    -------------- & ------------\\
     z_{1}(X),\mbox{NDSolve} & -0.71009\,78273\,59729\\
   -------------- & ------------\\
   y_{2}(X),\mbox{NDSolve} & -0.06656\,18185\,10763\,63\\
       -------------- & ------------\\
    z_{2}(X),\mbox{NDSolve} &  0.24186\,12563\,97462\,12\\
         -------------- & ------------   \end{array} \end{equation}
     \subsection{n=2 case}\label{n2case}
     In this case we first mention that recurrecy relation for coefficients
     $a_{i}$, may be obtained in the more simple form than in (\ref{aaii}).
     This relation, according to \cite {Seid79a}, is: \begin{equation}\label{n2ai}
a_{i+1}=-{1\over (2\,i+2)\,(2\,i+3)}\,
\sum_{k=0}^{k=i}\,a_{k}\,a_{i-k},\quad i\geq 0,\quad a_{0}=1. \end{equation}
 From equation (\ref{n2ai}) it follows that the series is sign-alternating,
 the result not evident from (\ref{aaii}). In general, it can be shown, that
 coefficients $a_{i}$ of serial solution for LEE of {\it integer} index $n$ is
sign-alternating. \\ We present results of numerical solution of the system
(\ref{sys4a}-\ref{sys4c}). \begin{equation}\label{2sol} \begin{array}{cl}
 -------------- & ------------\\  \,n=2 &  \\ -------------- & ------------\\
{X},\mbox{NDSolve}& 4.35287\,45959\,46124 \\
{X},{\mbox{\cite{Jabb84}}}& 4.35287\,45959 \\
{X},{\mbox{\cite{Hord90}}}& 4.35287\,5\\ -------------- & ------------\\
    y_{0}(X),\mbox{NDSolve} &   3.891892651074 \cdot 10^{-17} \\
  y_{0}(X),\mbox{Analytic} &  \mbox{ 0}\\ -------------- & ------------\\
  z_{0}(X), \mbox{NDSolve} & 2.41104\,60120\,96894 \\
   z_{0}(X),{\mbox{\cite{Hord90}}}  & 2.41104\,73856\\
-------------- & ------------\\
   y_{1}(X),\mbox{NDSolve} &  0.20990106855590432 \\
    -------------- & ------------\\
     z_{1}(X),\mbox{NDSolve} & -0.51626996767525199\\
   -------------- & ------------\\
   y_{2}(X),\mbox{NDSolve} & -0.049552916011399585\\
       -------------- & ------------\\
    z_{2}(X),\mbox{NDSolve} &  0.15512703899482025\\
         -------------- & ------------  \end{array} \end{equation}
     \subsection{n=5/2 case}\label{n25case}
We present results of numerical solution of the system (\ref{sys4a}-\ref{sys4c}).
 \begin{equation}\label{25sol} \begin{array}{cl} -------------- & ------------\\
  \,n=5/2 &  \\ -------------- & ------------\\
{X},\mbox{NDSolve}&         5.35527\,54590\,10824 \\
{X},{\mbox{\cite{Jabb84}}}& 5.35527\,54459\,0 \\
{X},{\mbox{\cite{Hord90}}}& 5.35527\,5\\-------------- & ------------\\
    y_{0}(X),\mbox{NDSolve} &  -3.385331 \cdot 10^{-15} \\
  y_{0}(X),\mbox{Analytic} & \mbox{ 0}\\  -------------- & ------------\\
  z_{0}(X), \mbox{NDSolve} & 2.18719\,95655\,17079 \\
   z_{0}(X),{\mbox{\cite{Hord90}}}  & 2.18719\,90907\\
-------------- & ------------\\  y_{1}(X),\mbox{NDSolve} &  0.18557608273853998 \\
    -------------- & ------------\\  z_{1}(X),\mbox{NDSolve} &-0.38690411187876034\\
   -------------- & ------------\\y_{2}(X),\mbox{NDSolve} & -0.03834028384550369\\
       -------------- & ------------\\  z_{2}(X),\mbox{NDSolve} & 0.10810493285592643\\
         -------------- & ------------  \end{array} \end{equation}
    \subsection{n=3 case}\label{n3case} \begin{equation}\label{3sol} \begin{array}{cl}
 -------------- & ------------\\ \,n=3 &  \\ -------------- & ------------\\
{X},\mbox{NDSolve}&       6.89684\,86193\,7696 \\
{X},{\mbox{\cite{Jabb84}}}&6.89684\,86194 \\ {X},{\mbox{\cite{Hord90}}}&6.89684\,9\\
-------------- & ------------\\
    y_{0}(X),\mbox{NDSolve} &  1.7861333071518752 \cdot 10^{-17} \\
  y_{0}(X),\mbox{Analytic} &  \mbox{ 0}\\   -------------- & ------------\\
  z_{0}(X), \mbox{NDSolve} & 2.01823\,59509\,66228\,7 \\
   z_{0}(X),{\mbox{\cite{Hord90}}}  & 2.01823\,62876\\
-------------- & ------------\\
   y_{1}(X),\mbox{NDSolve} & 0.16547\,68294\,39454\,83 \\
    -------------- & ------------\\
     z_{1}(X),\mbox{NDSolve} &-0.29335\,99255\,39480\,26\\
   -------------- & ------------\\
   y_{2}(X),\mbox{NDSolve} & -0.03053\,51512\,78657\,177\\
       -------------- & ------------\\
    z_{2}(X),\mbox{NDSolve} & 0.08152\,75811\,87124\,82\\
         -------------- & ------------      \end{array} \end{equation}
    \subsection{n=7/2 case}\label{n72case}
 \begin{equation}\label{72sol} \begin{array}{cl}
 -------------- & ------------\\ \,n=7/2 &  \\ -------------- & ------------\\
{X},\mbox{NDSolve}&       9.53580\,53442\,4479 \\
{X},{\mbox{\cite{Jabb84}}}&  9.53580\,53443\\
{X},{\mbox{\cite{Hord90}}}&  9.535805\\ -------------- & ------------\\
    y_{0}(X),\mbox{NDSolve} &  1.2322310495888572 \cdot 10^{-15} \\
  y_{0}(X),\mbox{Analytic} & \mbox{    0}\\
  -------------- & ------------\\
  z_{0}(X), \mbox{NDSolve} & 1.89055\,70934\,43116\,4 \\
   z_{0}(X),{\mbox{\cite{Hord90}}}  & 1.89055\,65987\\
-------------- & ------------\\
   y_{1}(X),\mbox{NDSolve} & 0.1481615824132385 \\
    -------------- & ------------\\
     z_{1}(X),\mbox{NDSolve} &-0.21964831524439676\\
   -------------- & ------------\\
   y_{2}(X),\mbox{NDSolve} & -0.024934678004649848\\
       -------------- & ------------\\
    z_{2}(X),\mbox{NDSolve} & 0.06771308534193725\\
         -------------- & ------------  \end{array} \end{equation}
         \subsection{n=4 case}\label{n4case}
 \begin{equation}\label{4sol} \begin{array}{cl}
 -------------- & ------------\\ \,n=4 &  \\ -------------- & ------------\\
{X},\mbox{NDSolve}&    14.97154\,63488\,38093 \\
{X},{\mbox{\cite{Jabb84}}}& 14.97154\,63496\\
{X},{\mbox{\cite{Hord90}}}& 14.97155\\ -------------- & ------------\\
    y_{0}(X),\mbox{NDSolve} &  9.088953198578 \cdot 10^{-18} \\
  y_{0}(X),\mbox{Analytic} &  \mbox{    0}\\  -------------- & ------------\\
  z_{0}(X), \mbox{NDSolve} & 1.79722\,99144\,3925 \\
   z_{0}(X),{\mbox{\cite{Hord90}}}  & 1.79723\,08344\\
-------------- & ------------\\
 y_{0}^{''}(X),\mbox{NDSolve} &  0.00107\,11089\,71589\,365 \\
-------------- & ------------\\
   y_{1}(X),\mbox{NDSolve} & 0.13250\,00877\,72273\,87 \\
    -------------- & ------------\\
     z_{1}(X),\mbox{NDSolve} &-0.15399\,30629\,85788\,74\\
   -------------- & ------------\\
   y_{2}(X),\mbox{NDSolve} & -0.02097\,11409\,23665\,716\\
       -------------- & ------------\\
    z_{2}(X),\mbox{NDSolve} & 0.06552\,46752\,25231\,59 \\
         -------------- & ------------ \end{array} \end{equation}
        \subsection{n=9/2 case}\label{n92case}
 \begin{equation}\label{92sol} \begin{array}{cl}
 -------------- & ------------\\  \,n=9/2 &  \\ -------------- & ------------\\
{X},\mbox{NDSolve}&      31.83646\,32446\,95228 \\
{X},{\mbox{\cite{Jabb84}}}&  31.83646\,32485\\
{X},{\mbox{\cite{Hord90}}}&  31.83646\\ -------------- & ------------\\
    y_{0}(X),\mbox{NDSolve} &  -1.6287136623021175 \cdot 10^{-15} \\
  y_{0}(X),\mbox{Analytic} &  \mbox{    0}\\  -------------- & ------------\\
  z_{0}(X), \mbox{NDSolve} & 1.73779\,88676\,66032\,3 \\
   z_{0}(X),{\mbox{\cite{Hord90}}}  & 1.73779\,86022\\
-------------- & ------------\\
   y_{1}(X),\mbox{NDSolve} & 0.11719722297815073 \\
    -------------- & ------------\\
     z_{1}(X),\mbox{NDSolve} &-0.07993661318489896\\
   -------------- & ------------\\
   y_{2}(X),\mbox{NDSolve} & -0.018685417777349604\\
       -------------- & ------------\\
    z_{2}(X),\mbox{NDSolve} & 0.08780938436898907\\
         -------------- & ------------    \end{array} \end{equation}
\section{Global parameters} \subsection{Radius}
Radius of the polytrope is proportional to $X$, the first zero of $y(x)$.
In our second order approximation in $\delta$, we write:
\begin{equation}\label{X}
 X=X_{0}+\delta\,X_{1}+\delta^{2}\,X_{2},\quad y(X)=0,\quad
 y(X)=y_{0}(X)+\delta \,y_{1}(X) +\delta^{2}\,y_{2}(X),
\end{equation} from where we get: \begin{equation}\label{X12}
 X_{1}={y_{1}(X_{0})\over -y_{0}^{'}(X_{0})};\quad
 X_{2}={y_{2}(X_{0})+X_{1}\,y_{1}^{'}(X_{0})+1/2\, X_{1}^{2}\,y_{0}^{''}(X_{0})\over
 -y_{0}^{'}(X_{0})}. \end{equation}
Let me remind the meaning of $X_{0},\,X_{1}$, and $X_{2}$: for each $n$,
$X_{0}$ is the dimensional radius, while $X_{1}$, and $X_{2}$ define the
derivatives of $X(n)$ by $n$: $X_{0}=X(n)$, $X_{1}=d\,X(n)/\,d\,n$, and
$2\,X_{2}=d^{2}\,X(n)/\,d\,n^{2}$.  \subsection{Mass}
Mass of the polytrope is proportional to $\mu=-X^{2}\,y^{'}(X)$.
In our second order approximation in $\delta$, we write:\begin{equation}\label{M}
 \mu=\mu_{0}+\delta\,\mu_{1}+\delta^{2}\,\mu_{2},\quad \end{equation}
 from where we get:
\begin{equation}\label{M1} \mu_{0}=-X_{0}^{2}\,y_{0}^{'}(X_{0}),\quad
 \mu_{1}=-X_{0}^{2}\,[X_{1}\,y_{0}^{''}(X_{0}) +y_{1}^{'}(X_{0})]
 - 2 X_{0}\,X_{1}\,y_{0}^{'}(X_{0}). \end{equation}
\begin{equation}\label{M2}\begin{array}{l}
 \mu_{2}=-X_{0}^{2}\,\left[X_{2}\,y_{0}^{''}(X_{0})+
 {1\over 2}\,X_{1}^{2}\,y_{0}^{'''}(X_{0})+ X_{1}\,y_{1}^{''}(X_{0})+
 y_{2}^{'}(X_{0})\right]\\-2 X_{0}\,X_{1}\,\left(X_{1}\,y_{0}^{''}(X_{0})
 +y_{1}^{'}(X_{0})\right)-y_{0}^{'}(X_{0})\,(X_{1}^{2}+ 2\,X_{0}\,X_{2}).\end{array}
\end{equation} Let me remind the meaning of $\mu_{0},\,\mu_{1}$, and $\mu_{2}$: for each $n$,
$\mu_{0}$ is the dimensional mass, while $\mu_{1}$, and $\mu_{2}$ define the
derivatives of $\mu(n)$ by $n$: $\mu_{0}=\mu(n)$, $\mu_{1}=d\,\mu(n)/\,d\,n$, and
$2\,\mu_{2}=d^{2}\,\mu(n)/\,d\,n^{2}$. \newpage \subsubsection{n=0}
\begin{equation}\label{n0}\begin{array}{l} n=0, \\ X_0 =  \surd {6},\\
  y_0^{'}(X_0) =   -0.81649\,65809\,277260,\\
  y_0^{''}(X_0) =  -0.33333\,33333\,33333,\\  y_0^{'''}(X_0) =  0,\\
  y_1^{}(X_0) =   0.43925\,53889\,064479,\\
  y_1^{'}(X_0) =    1.04541\,96097\,934910777,\\
  y_1^{''}(X_0) =  70.10101\,52396\,6198, \\
  y_2^{}(X_0) =  -0.41351\,38937\,12500\,52051\,096,\\
   y_2^{'}(X_0) = 4.33084\,54146\,19353 \cdot 10^{11},\\
    X_1 =         0.53797\,57847\,94289\,96\,7933774,\\
     X_2 =        0.12328\,30846\,88831\,89\,0159441,\\
  X(n=.1) =       2.50452\,01521\,09495,(*  2.5045449630\, \mbox{Jabbar} *)\\
  X(n=-.1) =  -,\\   \mu_0 =  4.89897\,94855\,66356\,196,\\
  \mu_1 =  -3.04466\,29499\,952067,\\
  \mu_2 =  -2.59850\,72489\,991943\cdot 10^{12},\\
  \mu(n=.1) =  4.59451\,31905,\, (* \mbox{Horedt} *)\\
  \mu(n=-.1) =  -.\end{array}\end{equation}\newpage\subsubsection{n=1/2}
\begin{equation}\label{n12}\begin{array}{l} n = 1/2, \\ X_0 = 2.75269805406499,\\
    y_0 ^{'}(X_0) = -0.49999708294422937513122874,\\
    y_0^{''}(X_0) =  0.3632777637699180575486,\\
    y_0^{'''}(X_0) = 3.7771115383\cdot 10^6,\\
    y_1(X_0) =  0.34123204013164497867521625893, \\
    y_1^{'}(X_0) =  0.215076348248895845791,\\
    y_1^{''}(X_0) = -2.723045849967156701\cdot 10^6,\\
    y_2(X_0) = -0.146821475444243995570334,\\
    y_2^{'}(X_0) = 929115.93812128759,\\    X_1 = 0.6824680618580861779508,\\
     X_2 = 0.169124728807637034976,\\
      X(n=.6) =  2.822636107538865,(*  2.8226750739,\, \mbox{Jabbar} *)\\
    X(n=.4) =  2.686142495167248,(*  2.6861053263,\, \mbox{Jabbar} *)\\
    \mu_0 =  3.7886511848840057435396,\\ \mu_1 = -1.6297077071514047746998,\\
    \mu_2 =  0,\\ \mu(n=.6) =  3.6256804141688654,\,\mbox{(only 1st appr!)} \\
     (*\, 0.4560739*2.822675^2=3.63376\,61327, \mbox{Horedt}\\
     \mu(n=.4) =    3.9516219555991463,\,\mbox{only 1st appr!}\\
     (*\, 0.5489336*2.686105^2=3.96064\,37923, \mbox{Horedt}\, *).
\end{array}\end{equation} \newpage \subsubsection{n=1}
\begin{equation}\label{n1}\begin{array}{l} X_{0}=\pi,\\
y_{0}^{'}(X_{0}) = -0.31830988618379067039,\\
y_{0}^{''}(X_{0}) = 0.202642367284675542812372,\\
y_{0}^{'''}(X_{0}) = 0.12480067956938`3.2195,\\
y_{1}(X_{0}) = 0.2817914499022078201593,\\ y_{1}^{'}(X_{0}) = 0.104285289178956941658, \\
y_{1}^{''}(X_{0}) =-0.348181526960602809649062, \\
y_{2}(X_{0}) = -0.09455031889587342081743776,\\
y_{2}^{'}(X_{0})= 0.0822442339327594290586,\\ X_{1}=  0.88527394885719235129261239,\\
X_{2}=   0.2424591740663775292855118679,\\
X(n=1.1)= 3.232544640216176,\, (*\, 3.2326084072,\, \mbox{Jabbar} *)\\
X(n=0.9)= 3.0554898504447374,  (*\, 3.0554293447,\,  \mbox{Jabbar}\, *)\\
\mu_{0}=         3.1415926535897932271, \\ \mu_{1}= -1.02925454904951,\\
\mu_{2}= 0.4193306804206245,\\ \mu(n=1.1)= 3.0428605054890485,\\
  (* 0.2911738* 3.232608^2 = 3.042694721493137,\, \mbox{Horedt} *)\\
\mu(n=.9)= 3.2487114152989505,\\ (* 0.3055429*3.055429^2 =
3.24889082611291,\, \mbox{Horedt} *).\end{array}\end{equation}
\newpage \subsubsection{n=3/2}
\begin{equation}\label{n32}\begin{array}{l} n=3/2,\\ X_{0}=3.653753736219119,\\
y_0 ^{'}(X_0) =  -0.203301282638546737276897,\\
y_0 ^{''}(X_0) = 0.111283516797123609328,\\
y_0^{'''}(X_0) =  -0.09137193541041363,\\ y_1(X_0) =   0.24067091914095230177155721,\\
y_1^{'}(X_0) = 0.0531911817234202858,\\ y_1^{''}(X_0) = -0.02911591910998444970746,\\
y_2(X_0) = -0.066561818510763637968669,\\ y_2^{'}(X_0) = -0.018117061781318394264,\\
X_1 = 1.183814071497255432331730,\\ X_2 = 0.3658800713499040420899,\\
X(n = 1.6) = 3.7757939440823436,\, (*\, 3.77590\,47640,\, \mbox{Jabbar} *)\\
X(n = 1.4) = 3.5390311297828925,\, (*\, 3.\, (*\, 3.53892\,66160,\, \mbox{Jabbar} *)\\
\mu_0 = 2.7140551201086457355,\\ \mu_1 = -0.71009782735972906926,\\
\mu_2 = 0.24186141001555074558,\\ \mu(n = 1.6) = 2.64546\,39514\,728284,\\
(* \, 0.185544*3.775905^2 = 2.64538\,58927,\, \mbox{Horedt} *)
\mu(n = 1.4) = 2.78748\,35169\,44774,\\
(* 0.2225779*3.538927^2 = 2.78756\,657920,\, \mbox{Horedt} *)
.\end{array} \end{equation} \newpage \subsubsection{n=2}
\begin{equation}\label{n2}\begin{array}{l} n=2,\\ X_0 = 4.352874595946124,\\
 y_0 ^{'}(X_0) = -0.12724865113117523,\\  y_0^{''}(X_0) = 0.05846649074139796,\\
 y_0^{'''}(X_0) = -0.040295089683388775,\\  y_1(X_0) = 0.20990106855590432,\\
 y_1^{'}(X_0) = 0.027247367605845233,\\  y_1^{''}(X_0) = -0.012519252280422236,
 y_2(X_0) = -0.049552916011399585,\\  y_2^{'}(X_0) = -0.008187196082955055,\\
 X_1 = 1.6495347234724416,\\  X_2 = 0.5888879298143976,\\
 X(n=2.1) = 4.523716947591513,\,(*\, 4.5239262993,\,  \mbox{Jabbar}\, *)\\
 X(n=1.9) = 4.193810002897024,\,(*\, 4.1936146217,\,  \mbox{Jabbar}\, *)\\
 \mu_0 = 2.411046012096894,\\  \mu_1 = -0.5162699676752518,\\
 \mu_2 = 0.15512703899481983,\\  \mu(n=2.1) = 2.360970285719317,\\
 (*\, 0.1153592*4.523926^2=2.3609305957478286, \mbox{Horedt}\, *)\\
 \mu(n=1.9) = 2.4642242792543674,\\
(*\, 0.140123*4.193615^2=2.4642600755839923, \mbox{Horedt}\, *).\end{array}
\end{equation} \newpage \subsubsection{n=5/2}
\begin{equation}\label{n52}\begin{array}{l} n=5/2,\\ X_0 =5.3552754590108239,\\
 y_0 ^{'}(X_0) = -0.076264913479991799518,\\
 y_0^{''}(X_0) = 0.0284821627061845798710092,\\
 y_0^{'''}(X_0) = -0.0159555729247840799988,\\
 y_1 (X_0) = 0.1855760827385399754972,\\
 y_1^{'}(X_0) = 0.01349086250870339952973971,\\
 y_1^{''}(X_0) = -0.005038344941156511,\\  y_2(X_0) = -0.03834028384550369027000,\\
 y_2^{'}(X_0) = -0.003769483809799692567,\\  X_1 = 2.43330\,87690\,08517\,98883,\\
 X_2 = 1.03335\,16315\,07183\,66045,\\
 X(n=2.6) = 5.608939852226747,\,(*\, 5.60938\,27386,\,  \mbox{Jabbar}\, *)\\
 X(n=2.4) =  5.122278098425044,\,(*\,  5.122278,\,  \mbox{Jabbar}\, *)\\
 \mu_0 = 2.18719956551707901061,\\  \mu_1 = -0.3869041118787603560,\\
 \mu_2 = 0.1081049328559264189, \\  \mu(n=2.6) =  2.14959\,02036\,57762, \\
(* \, 0.06831578*5.609382^2 = 2.14956\,73869,\,\mbox{Horedt} \, *)\\
\mu(n=2.4) =  2.22697\,10260\,33514,\\
(* \, 0.08489109*5.121870^2 = 2.22699\,48490 ,\,\mbox{Horedt} \, *).
\end{array}\end{equation} \newpage \subsubsection{n=3}
\begin{equation}\label{n3}\begin{array}{l} n=3,\\
X_0 = 6.89684861937696,\\ y_0 ^{'}(X_0) = -0.04242975760445296,\\
y_0^{''}(X_0) =  0.012304100016127638, \\ y_0^{'''}(X_0) =  -0.005352053102589007,\\
y_1(X_0) = 0.16547682943945483,\\ y_1^{'}(X_0) =  0.006167361415567608,\\
y_1^{''}(X_0) =  -0.0017884578177459688,\\ y_2(X_0) = -0.030535151278657177,\\
y_2^{'}(X_0) =  -0.0017139698191338721,\\ X_1 = 3.9000182603467954,\\
X_2 =  2.0525978144309858,\\
X(n=3.1) = 7.30737\,64235,\,(*\,7.30848\,42924\,  \mbox{Jabbar}\, *)\\
X(n=2.9) = ", 6.52737277148659,(*\, 6.52637\,41261\,  \mbox{Jabbar}\, *)\\
\mu_0 = 2.0182359509662287,\\ \mu_1 = ", -0.29335992553948076,\\
\mu_2 = 0.08152758134658877,\\ \mu(n=3.1) = ", 1.98971\,52342\,25746,\\
(*\, 0.03725063*7.308484^2=1.98970\,28553, \mbox{Horedt}\, *)\\
\mu(n=2.9) =  2.0483872193336428,\\
(*\, 0.04809180*6.526374^2=2.04840\,08528, \mbox{Horedt}\, *).
\end{array} \end{equation} \newpage \subsubsection{n=7/2}
\begin{equation}\label{n72}\begin{array}{l} n=7/2,\\ X_0 = 9.53580534424479,\\
y_0 ^{'}(X_0) =-0.020790983939331436621,\\ y_0^{''}(X_0) = 0.00436061416708334168,\\
y_0^{'''}(X_0) = -0.00137186551413256,\\  y_1(X_0) = 0.148161582413238524348,\\
y_1^{'}(X_0) = 0.0024155338182516865,\\ y_1^{''}(X_0) = -0.00050662397795474,\\
 y_2(X_0) = -0.0249346780046498475,\\  y_2^{'}(X_0) = -0.0007446596956576678,\\
  X_1 = 7.12624197323115573338729,\\ X_2 = 4.9541783719117226,\\
X(n=3.6) = 10.297971325287023,\,(*\, 10.30159\,12890,\,  \mbox{Jabbar}\, *)\\
X(n=3.4) =  8.872722930640792,\,(*\, 8.8695802788,\,  \mbox{Jabbar}\, *)\\
\mu_0 = 1.890557093443116455,\\ \mu_1 = -0.219648315244396768,\\
\mu_2 = 0.067713085342,\\ \mu(n=3.6) = 1.86926\,93927\,72097,\\
(*\,0.01761417*10.30159^2 = 1.86926\,42743,  \mbox{Horedt}\, *)\\
\mu(n=3.4) =  1.91319\,90558\,20976,\\
(*\, 0.02431955*8.869580^2 = 1.91320\,56075 ,  \mbox{Horedt}\, *)\\
\end{array} \end{equation} \newpage \subsubsection{n=4}
\begin{equation}\label{n4}\begin{array}{l} n=4,\\ X_0 =  14.971546348838093,\\
y_0^{'}(X_0) = -0.008018078806403244,\\ y_0^{''}(X_0) =  0.0010711089715893653,\\
y_0^{'''}(X_0) =  -0.0002146289260907,\\ y_1(X_0) = 0.13250008777227387,\\
y_1^{'}(X_0) =  0.0006870175622715013,\\ y_1^{''}(X_0) =  -0.00009177643327735738,\\
y_2(X_0) = -0.020971140923665713,\\ y_2^{'}(X_0) =  -0.0002923287696798706,\\
X_1 = 16.525166560655304,\\  X_2 =  17.040461450049975,\\
X(n=4.1) =  16.7944\,67619,\,(*\, 16.81377\, 38705,  \mbox{Jabbar}\, *)\\
X(n=3.9) =  13.48943\,43072,\,(*\, 13.47384\,13948,  \mbox{Jabbar}\, *)\\
    \mu_0 =  1.7972299144392503,\\  \mu_1 =  -0.1539930629857884,\\
    \mu_2 =  0.06552467522560379,\\ \mu(n=4.1) =  1.78248\,58548\,92927,\\
   (*\, 0.006305183*16.81378^2 = 1.78249\,53973,  \mbox{Horedt}\, *)\\
    \mu(n=3.9) =   1.81328\,44674\,90085,\\
  (*\,  0.009988109*13.47384^2 = 1.81328\,48994, \mbox{Horedt}\, *).\\
\end{array} \end{equation} \newpage \subsubsection{n=9/2}
\begin{equation}\label{n45}\begin{array}{l} n=9/2,\\ X_0 =  31.836463244695223, \\
y_0 ^{'}(X_0) = -0.0017145489124289180377, \\ y_0^{''}(X_0) = 0.0001077097602991190, \\
y_0^{'''}(X_0) = -0.000010149660105577,\\ y_1(X_0) = 0.1171972229781507365218,\\
 y_1^{'}(X_0) = 0.000078867143804445547,\\
  y_1^{''}(X_0) = -4.95451666212243895495976\cdot 10^{-6},\\
  y_2(X_0) = -0.01868541777734960517998,\\   y_2^{'}(X_0) = -0.00008663458543571578,\\
    X_1 = 68.35455210905772991040946,\\    X_2 = 139.00687095347173451822,\\
     X(n=4.6) = 40.06198\,71651,\,(*\, 40.41322\,96343,  \mbox{Jabbar}\, *)\\
      X(n=4.4) = 26.39107\,67433,\,(*\,  26.15891\,96485,  \mbox{Jabbar}\, *)\\
      \mu_0 = 1.7377988676660324,\\    \mu_1 = -0.0799366131848989555,\\
       \mu_2 = 0.087809384368989,\\       \mu(n=4.6) =  1.7306833001912325,\\
(*\,  0.001059709*40.41324^2 = 1.73074\,84954, \mbox{Horedt}\, *),\\
\mu(n=4.4) =  1.7466706228282123,\\
(*\,  0.002552517*26.15892^2 = 1.74665\,95493, \mbox{Horedt}\, *).\\
\end{array} \end{equation} \newpage \section{Summary}
In this paper we present the exact numerical solutions for the internal structure
and global parameters of polytropic models of arbitrary index $0 <\,n < 5$.
The  perturbation method used here is
not rigorously founded by means of the theory of differential equations and there
are some problems about application of the method in the interval of argument
 where the perturbation function is of the same order or  even larger than
 the initial\,non-perturbed function.  The problem of justification is
 here similar to the problem of rotationally distorted polytropes which was
 already discussed in the astrophysical literature  (see e.g. \cite{ChLe62}).
 Still validity of any method in applicational sciences (as astrophysics is such
 relative to mathematics) may be checked with\,\,numerical calculations
 and (astro)-physical "common sense", and
 we showed in this paper that the perturbation method is applicable to the
 problem of the structure and the global parameters of the polytropic models.

\begin{thebibliography}{}
\bibitem[Abramowitz \& Stegun(1972)]{AbSt72}
Abramowitz, M. \& Stegun, I. A. (Eds.) 
Handbook of Mathematical Functions with Formulas, Graphs, and Mathematical Tables,
9th ed.\,new York: Dover,  1972
\bibitem[Caimmi(1987)]{Caim87} Caimmi, R.  1987, \apss, 140, 1
\bibitem[Chandrasekhar(1957)]{Chan57} Chandrasekhar, S. 1957,
 An Introduction to the Study of Stellar Structure, Chicago: Dover, 1957
\bibitem[Chandrasekhar \& Lebovitz (1962)]{ChLe62}
Chandrasekhar, S. \& Lebovitz,\,n. R. 1962, \apj, 136, 1082
  \bibitem[Christensen-Dalsgaard \& Mullan(1994)]{CrMu94}
 Christensen-Dalsgaard, J. \& Mullan, D. J. 1994, \mnras,  270, 921
\bibitem[Gradshteyn \& Ryzhik (1980)]{GrRy80} Gradshteyn, I. S. \& Ryzhik, I. M. 1980,
Tables of Integrals, Series, and Products,\,new York: Academic Press, 1980
\bibitem[Horedt(1990)]{Hord90} Horedt, G. P.  1990, \apj, 357, 560
\bibitem[Jabbar(1984)]{Jabb84} Jabbar, J. R.  1984, \apss, 100, 447
\bibitem[Jeans(1929)]{Jean29} Jeans, J. 1929,
 Astronomy and Cosmogony,\,new York: Dover, 1929
\bibitem[Kamke (1959)]{Kamk59} Kamke, E. 1959,
Differentialgleichunge, 6th ed., Leipzig, 1959 (2.5.)
\bibitem[Korn \& Korn (1968)]{KoKo68} Korn, G. A. \& Korn, T. M. 1968,
Mathematical Handbook, 2nd ed.,\,new York: McGraw-Hill, 1968 (10.2-7.c)
  \bibitem[Milne(1929)]{Miln29}
 Milne, E. A.  1929, \mnras,  89, 739
\bibitem[Horedt(1987)]{Hord87} Horedt, G. P.  1987,
Astron. Ap., 172, 359
\bibitem[Medvedev \& Rybicki(2001)]{MoRy01} Medvedev, M. V. \&  Rybicki, G.
2001, \apj, 555, 863
\bibitem[Seidov (1978a)]{Seid78a} Seidov, Z. F. 1978,
 Sov. Astron. Lett., 4(3), 144
 \bibitem[Seidov (1978b)]{Seid78b} Seidov, Z. F. 1978,
in Sources of Gravitational Radiation, Proceed. Batelle Seattle Workshop,
Smarr L.L. ed.,\,new York: CUP (pp. 480-482)
 \bibitem[Seidov (1979a)]{Seid79a} Seidov, Z. F. 1979,
Dokl. AN AzSSR, 35,\,no. 1, 21
 \bibitem[Seidov (1979b)]{Seid79b} Seidov, Z. F. 1979,
in IAU Coll.\,no. 53, Rochecter, 1979, (pp. 478-482)
Smarr L.L. ed.,\,new York: CUP (pp. 480-482)
 \bibitem[Seidov (2004)]{Seid04}
  Seidov, Z. F. 2004, preprint (astro-ph/0401359)
\bibitem[Seidov \& Kuzakhmedov(1977)]{SK77}
Seidov, Z. F., \& Kuzakhmedov, R. Kh. 1977,
 Sov. Astron., 21, 399 (SK77)
\bibitem[ Seidov \& Kuzakhmedov(1978)]{SK78}
Seidov, Z. F., \& Kuzakhmedov, R. Kh. 1978,
 Sov. Astron.,  22, 711 (SK78)
\bibitem[Wolfram (1999)]{Wolf99}  Wolfram, S. 1999,
The Mathematica Book, 4th ed., Cambridge: Wolfram Media/CUP, 1999
\end{thebibliography}
\end{document}